\documentclass[preprint,nofootinbib]{revtex4}%
\usepackage{amssymb}
\usepackage{amsfonts}
\usepackage{amsmath}
\usepackage{graphicx}%
\newcommand{\beq}{\begin{equation}}
\newcommand{\eeq}{\end{equation}}
\newcommand{\bea}{\begin{eqnarray}}
\newcommand{\eea}{\end{eqnarray}}
\begin{document}
\title{Static solutions with nontrivial boundaries\\for the Einstein-Gauss-Bonnet theory in vacuum}
\author{Gustavo Dotti$^{1,5}$, Julio Oliva$^{2}$, and Ricardo Troncoso$^{3,4}$}
\email{gdotti@famaf.unc.edu.ar, julio.oliva@docentes.uach.cl, troncoso@cecs.cl}
\affiliation{$^{1}$FaMAF, Universidad Nacional de C\'{o}rdoba, Ciudad Universitaria, (5000)
C\'{o}rdoba, Argentina. }
\affiliation{$^{2}$Instituto de F\'{\i}sica, Facultad de Ciencias, Universidad Austral de Chile,}
\affiliation{$^{3}$Centro de Estudios Cient\'{\i}ficos (CECS), Casilla 1469, Valdivia, Chile.}
\affiliation{$^{4}$Centro de Ingenier\'{\i}a de la Innovaci\'{o}n del CECS (CIN), Valdivia, Chile.}
\affiliation{$^{5}$Instituto de F\'{\i}sica Enrique Gaviola, CONICET.}
\preprint{CECS-PHY-09/09}

\begin{abstract}
The classification of certain class of static solutions for the
Einstein-Gauss-Bonnet theory in vacuum is performed in $d\geq5$ dimensions.
The class of metrics under consideration is such that the spacelike section is
a warped product of the real line and an arbitrary base manifold. It is shown
that for a generic value of the Gauss-Bonnet coupling, the base manifold must
be necessarily Einstein, with an additional restriction on its Weyl tensor for
$d>5$. The boundary admits a wider class of geometries only in the special
case when the Gauss-Bonnet coupling is such that the theory admits a unique
maximally symmetric solution. The additional freedom in the boundary metric
enlarges the class of allowed geometries in the bulk, which are classified
within three main branches, containing new black holes and wormholes in vacuum.

\end{abstract}
\maketitle

\section{Introduction}

\label{intro}

The asymptotic properties of spacetime play a crucial role for a suitable
definition of energy in gravitation, which has been a subtle issue since the
early days of General Relativity (see, e.g. \cite{Julia-Silva}). Nowadays,
understanding the asymptotic structure of spacetime becomes a fundamental
problem by itself. In the case of negative cosmological constant, the
asymptotic behaviour of gravity is particularly interesting, and a renewed
interest has been raised in view of the AdS/CFT correspondence, which is a
conjectured duality between gravity on asymptotically AdS spacetimes and
conformal field theory (for a review see e.g., \cite{MAGOO}). In this context,
it is natural wondering about the possible freedom in the choice of the metric
at the boundary, where the dual theory is defined. As a simple example, one
can consider the following class of $d$-dimensional static metrics in bulk%
\begin{equation}
ds^{2}=-f^{2}\left(  r\right)  dt^{2}+\frac{dr^{2}}{g^{2}\left(  r\right)
}+r^{2}d\Sigma_{(d-2)}^{2}\ ,\label{Ansatz}%
\end{equation}
where
\begin{equation}
d\Sigma_{(d-2)}^{2}=\tilde{g}_{ij}(x)dx^{i}dx^{j}\ ,\label{base}%
\end{equation}
is the line element\footnote{Here $x^{i}$ correspond to local
\textquotedblleft angular\textquotedblright\ coordinates, and hereafter a
tilde is used on geometrical objects intrinsically defined on $\Sigma_{(d-2)}%
$.} of the \textquotedblleft base\textquotedblright\ manifold $\Sigma_{(d-2)}$
of $d-2$ dimensions.

The Einstein equations with cosmological constant $\Lambda$ in vacuum are then
solved for
\begin{equation}
f^{2}=g^{2}=\gamma-\frac{\mu}{r^{d-3}}-\frac{2\Lambda}{(d-1)(d-2)}r^{2}\ ,
\label{f Einstein}%
\end{equation}
provided the geometry of $\Sigma_{(d-2)}$ is restricted to be that of an
Einstein manifold, fulfilling%
\begin{equation}
\tilde{R}_{\ j}^{i}=(d-3)\gamma\delta_{j}^{i}\ , \label{Einstein Base}%
\end{equation}
where the constant $\gamma$ can be normalized to $\pm1$ or zero
\cite{Birmingham, Gibbons-Hartnoll, Gibbons-Hartnoll-Pope}. Thus, if the
cosmological constant is nonnegative, solutions of the form (\ref{Ansatz}),
with (\ref{f Einstein}) and (\ref{Einstein Base}) describe black holes only
for $\gamma=1$ and $\mu>0$, otherwise they possess naked singularities.
Remarkably, for the asymptotically AdS case, the solution describes black
holes for any value of $\gamma$ provided $\mu$ is bounded from below
\cite{Mann1,Lemos}, widening the possibilities in order to define a dual
theory at the boundary, whose metric is of the form $R\times\Sigma_{(d-2)}$.

In dimensions greater than four, General Relativity (GR) is not the only
option to describe gravity. Indeed, a natural and conservative generalization
of GR, being the most general theory of gravity leading to second order field
equations for the metric is described by the Lovelock action, which possesses
nonlinear terms in the curvature in a precise combination \cite{Lovelock}. The
simplest case corresponds to the so-called Einstein-Gauss-Bonnet (EGB) theory,
whose action is quadratic in the curvature and it is given by%
\begin{equation}
I=\int\sqrt{-g}d^{d}x\ \left[  c_{1}R-2c_{0}+\frac{c_{2}}{2}\left(
R^{\alpha\beta\mu\nu}R_{\alpha\beta\mu\nu}-4R^{\mu\nu}R_{\mu\nu}+R^{2}\right)
\right]  \ ,\label{Action EGB}%
\end{equation}
so that apart from the Newton and cosmological constants, the theory possesses
and additional coupling $c_{2}$ associated with the quadratic terms. The field
equations read%
\begin{equation}
c_{2}\ H_{\mu\nu}+c_{1}\ G_{\mu\nu}+c_{0}\ g_{\mu\nu}=0\ ,\label{egb}%
\end{equation}
where $G_{\mu\nu}$ is the Einstein tensor, and%
\begin{equation}
H_{\mu\nu}:=RR_{\mu\nu}-2R_{\mu\rho}R_{\ \nu}^{\rho}-2R_{\ \rho}^{\delta
}R_{\ \mu\delta\nu}^{\rho}+R_{\mu\rho\delta\gamma}R_{\nu}^{\ \rho\delta\gamma
}+\frac{1}{d-4}Hg_{\mu\nu}\ ,\label{egb2}%
\end{equation}
with
\begin{equation}
H:=H_{\ \mu}^{\mu}=\frac{\left(  4-d\right)  }{4}\left(  R^{\alpha\beta\mu\nu
}R_{\alpha\beta\mu\nu}-4R^{\mu\nu}R_{\mu\nu}+R^{2}\right)  \ ,\label{s}%
\end{equation}
identically vanishes in $d<5$ dimensions.

In terms of the vielbein $e^{a}=e_{\mu}^{a}dx^{\mu}$ and the curvature
$2$-form $R^{ab}=\frac{1}{2}R_{\ \mu\nu}^{ab}dx^{\mu}dx^{\nu}$, the field
equations read%

\begin{equation}
\mathcal{E}_{a}:=\epsilon_{ab_{1}...b_{d-1}}\!\left[  a_{2}R^{b_{1}b_{2}%
}R^{b_{3}b_{4}}\!+2a_{1}R^{b_{1}b_{2}}e^{b_{3}}e^{b_{4}}\!+a_{0}e^{b_{1}%
}e^{b_{2}}e^{b_{3}}e^{b_{4}}\right]  \ e^{b_{5}}\!...e^{b_{d-1}}%
=0\ ,\label{feq}%
\end{equation}
where wedge product between forms is understood. The relation between the
constants $\alpha_{j}$ in (\ref{feq}) and $c_{j}$ in (\ref{egb}) is%
\begin{equation}
c_{0}=\frac{a_{0}}{2}\left(  d-1\right)  !,\;\;\;c_{1}=-2\left(  d-3\right)
!a_{1},\;\;\;c_{2}=-2\left(  d-5\right)  !a_{2}\ .\label{constants}%
\end{equation}

Generically, the field equations of the EGB theory admit two different
maximally symmetric solutions -(A)dS or Minkowski-, fulfilling\footnote{Here
$\delta_{\gamma\delta}^{\alpha\beta}:=\delta_{\gamma}^{\alpha}\delta_{\delta
}^{\beta}-\delta_{\delta}^{\alpha}\delta_{\gamma}^{\beta}$.}%
\begin{equation}
R_{\ \gamma\delta}^{\alpha\beta}=\lambda\delta_{\gamma\delta}^{\alpha\beta
}\ ,\label{ccs}%
\end{equation}
with two different radii, determined by%
\begin{equation}
\lambda_{\pm}=\frac{a_{1}}{a_{2}}\left(  -1\pm\sqrt{1-\frac{a_{2}a_{0}}%
{a_{1}^{2}}}\;\right)  \ .\label{cc}%
\end{equation}
In the limit of vanishing Gauss-Bonnet coupling, $a_{2}\rightarrow0$, the
branch with negative sign in (\ref{cc}) diverges, whereas the other gives the
expected GR limit, i.e., $\lambda_{+}=-\frac{a_{0}}{2a_{1}}$.

If the Gauss-Bonnet coupling is such that the square root in (\ref{cc})
vanishes, i.e.,%
\begin{equation}
a_{2}=\frac{a_{1}^{2}}{a_{0}}\ ,\label{sc}%
\end{equation}
the EGB theory admits a unique maximally symmetric vacuum. This case is
naturally singled out as \textquotedblleft special\textquotedblright,\ since
the theory admits solutions with a relaxed asymptotic behavior as compared
with the standard one of GR \cite{BHS}.

\bigskip

Concerning the possible freedom in the choice of boundary metrics for the
class of static spacetimes of the form (\ref{Ansatz}), it can be seen that the
presence of quadratic terms in the action generically leads to strong
restrictions on geometry of the boundary, determined by $\Sigma_{d-2}$, since
it has to be Einstein with supplementary conditions involving its Weyl tensor
\cite{Dotti-Gleiser, DOT5d, JP}. Nevertheless, in the special case (\ref{sc}),
the EGB theory admits a wider class of boundary metrics, such that
$\Sigma_{d-2}$ is not necessarily Einstein. The additional freedom in the
boundary metric enlarges the class of allowed geometries in the bulk, which
are classified within three main branches, containing new black holes and
wormholes in vacuum.

\bigskip

The class of static metrics of the form (\ref{Ansatz}) with (\ref{base}),
solves the field equations of the EGB theory in $d$ dimensions according to
the following scheme:

\subsection{$d=5$ dimensions}

\noindent$\mathbf{\circ}$\textbf{\ (i) Generic class:} For an arbitrary value
of the Gauss-Bonnet coupling $a_{2}$, the metric (\ref{Ansatz}) solves the EGB
field equations provided the base manifold $\Sigma_{3}$ is necessarily of
constant curvature $\gamma$ (normalized to $\pm1,0$), i.e.,%
\begin{equation}
\tilde{R}_{\ kl}^{ij}=\gamma\delta_{kl}^{ij}\ ,
\end{equation}
and
\begin{equation}
f^{2}=g^{2}\left(  r\right)  =\gamma+\frac{a_{1}}{a_{2}}r^{2}\left[  1\pm
\sqrt{\left(  1-\frac{a_{2}a_{0}}{a_{1}^{2}}\right)  +\frac{\mu}{r^{4}}%
}\;\right]  \ , \label{gcuadradogenerico}%
\end{equation}
where $\mu$ is an integration constant.\newline

\noindent$\mathbf{\circ}$\textbf{\ (ii) Special class:} In the special case
where the Gauss-Bonnet coupling is given by (\ref{sc}), The bulk geometries
split into three main branches according to the geometry of $\Sigma_{3}%
$:\newline

\noindent$\cdot$\textit{\ (ii.a) Black holes:}

For an arbitrary base manifold, i.e.,%
\begin{equation}
\Sigma_{3}:\text{ \textrm{arbitrary\ ,}}%
\end{equation}
the metric (\ref{Ansatz}) solves the field equations provided%
\begin{equation}
f^{2}=g^{2}=\sigma r^{2}-\mu~,\;\;\sigma:=\frac{a_{0}}{a_{1}}%
\ ,\label{gbhspecial1}%
\end{equation}
where $\mu$ is an integration constant.\newline

\noindent$\cdot$\textit{\ (ii.b1) Wormholes:}

For base manifolds $\Sigma_{3}$ of constant nonvanishing Ricci scalar,%
\begin{equation}
\tilde{R}=6\gamma\ ,
\end{equation}
the metric (\ref{Ansatz}) with
\begin{align}
f^{2}(r) &  =\left(  \sqrt{\sigma}r+a\sqrt{\sigma r^{2}+\gamma}\right)
^{2}~,\label{fcuaworm}\\
g^{2}\left(  r\right)   &  =\sigma r^{2}+\gamma~,\label{gcuaesp}%
\end{align}
is a solution of the field equations, where $a$ is an integration
constant.\newline

\noindent$\cdot$\textit{\ (ii.b2) Spacetime horns:}

If the base manifold $\Sigma_{3}$ has \emph{vanishing} Ricci scalar, i.e.,
\begin{equation}
\tilde{R}=0\ ,
\end{equation}
the solution is given by%
\begin{align}
f^{2}(r) &  =\left(  a\sqrt{\sigma}r+\frac{1}{\sqrt{\sigma}r}\right)
^{2}~,\label{f-horn}\\
g^{2}\left(  r\right)   &  =\sigma r^{2}~,\label{grr-horn}%
\end{align}
with $a$ an integration constant.\newline

\noindent$\cdot$\textit{\ (iii) Degeneracy:}

If $\Sigma_{3}$ is of constant curvature, i.e.,%
\begin{equation}
\tilde{R}_{\ kl}^{ij}=\gamma\delta_{kl}^{ij}\ ,
\end{equation}
then,%
\begin{align}
g^{2}\left(  r\right)   &  =\sigma r^{2}+\gamma\ ,\\
f^{2}\left(  r\right)   &  :\text{\textrm{an arbitrary function.}}%
\end{align}

\subsection{$d=6$ dimensions}

\noindent$\mathbf{\circ}$\textbf{\ (i) Generic class:} For arbitrary values of
the Gauss-Bonnet coupling the metric (\ref{Ansatz}) solves the EGB field
equations provided the base manifold $\Sigma_{4}$ is Einstein, i.e.,
\begin{equation}
\tilde{R}_{~j}^{i}=3\gamma\delta_{j}^{i}\ ,\label{einstein6}%
\end{equation}
(with $\gamma$ normalized to $\pm1,0$) with the following (scalar) condition:
\begin{equation}
\tilde{R}_{\ kl}^{ij}\tilde{R}_{~ij}^{kl}-4\tilde{R}_{ij}\tilde{R}^{ij}%
+\tilde{R}^{2}-24\xi=0\ ,\label{E4propxi}%
\end{equation}
and%
\begin{equation}
f^{2}\left(  r\right)  =g^{2}\left(  r\right)  =\gamma+\frac{a_{1}}{a_{2}%
}r^{2}\left[  1\pm\sqrt{\left(  1-\frac{a_{2}a_{0}}{a_{1}^{2}}\right)
+\frac{\mu}{r^{5}}+\frac{a_{2}^{2}}{a_{1}^{2}}\frac{\left(  \gamma^{2}%
-\xi\right)  }{r^{4}}}\right]  \ ,\label{generic6}%
\end{equation}
where $\xi$ and $\mu$ are integration constants.\newline

\noindent$\mathbf{\circ}$\textbf{\ (ii) Special class:} In the special case in
which the Gauss-Bonnet coupling is given by (\ref{sc}), the solutions splits
into three main branches according to the geometry of $\Sigma_{4}$:\newline

\noindent$\cdot$\textit{\ (ii.a.1) Black holes:}

The base manifold $\Sigma_{4}$ has the same restrictions as in the generic
case, i.e.,
\begin{equation}
\tilde{R}_{~j}^{i}=3\gamma\delta_{j}^{i}\ ,
\end{equation}
and%
\begin{equation}
\tilde{R}_{\ kl}^{ij}\tilde{R}_{~ij}^{kl}-4\tilde{R}_{ij}\tilde{R}^{ij}%
+\tilde{R}^{2}-24\xi=0\ ,
\end{equation}
with $f^{2}$ and $g^{2}$ given by%
\begin{equation}
f^{2}\left(  r\right)  =g^{2}\left(  r\right)  =\gamma+\frac{a_{1}}{a_{2}%
}r^{2}\left[  1\pm\sqrt{\frac{\mu}{r^{5}}+\frac{a_{2}^{2}}{a_{1}^{2}}%
\frac{\left(  \gamma^{2}-\xi\right)  }{r^{4}}}\right]  \ ,
\end{equation}
possessing a slower fall off at infinity as compared with (\ref{generic6}).

\bigskip

For the remaining branches, the base manifold $\Sigma_{4}$ \emph{is no longer
restricted to be Einstein}, but instead fulfills the following scalar
condition:%
\begin{equation}
\tilde{R}_{\ kl}^{ij}\tilde{R}_{~ij}^{kl}-4\tilde{R}_{ij}\tilde{R}^{ij}%
+\tilde{R}^{2}-4\gamma\tilde{R}+24\gamma^{2}=0\ ,\label{paratodas6}%
\end{equation}
and $g^{2}$ is given by%
\begin{equation}
g^{2}(r)=\sigma r^{2}+\gamma,\hspace{1cm}\sigma:=\frac{a_{0}}{a_{1}%
}\ .\label{forall}%
\end{equation}
\newline\noindent$\cdot$\textit{\ (ii.a.2) Special black holes:}

The base manifold is such that
\[
\Sigma_{4}:\text{\textrm{ no additional restriction besides (\ref{paratodas6}%
)\ ,}}%
\]
and
\begin{equation}
f^{2}\left(  r\right)  =g^{2}\left(  r\right)  =\sigma r^{2}+\gamma
\ .\label{ii.a.2.6}%
\end{equation}
\newline

\noindent$\cdot$\textit{\ (ii.b1) Wormholes:}

The base manifold $\Sigma_{4}$, besides (\ref{paratodas6}), has a nonvanishing
constant Ricci scalar,%
\[
\tilde{R}=12\gamma\ ,
\]
where $\gamma$ is rescaled to $\pm1$, and the metric is given by
\[
g^{2}(r)=\sigma r^{2}+\gamma\ ,
\]
and
\begin{equation}
f^{2}(r)=\left\{
\begin{array}
[c]{ccc}%
\left(  a\sqrt{\sigma r^{2}-1}+1-\sqrt{\sigma r^{2}-1}\tan^{-1}\left(
\frac{1}{\sqrt{\sigma r^{2}-1}}\right)  \right)  ^{2} & : & \gamma=-1\\
\left(  a\sqrt{\sigma r^{2}+1}+1-\sqrt{\sigma r^{2}+1}\tanh^{-1}\left(
\frac{1}{\sqrt{\sigma r^{2}+1}}\right)  \right)  ^{2} & : & \gamma=1
\end{array}
\right.  ~,\label{fcuaworm6}%
\end{equation}
with $a$ an integration constant.\newline\noindent$\cdot$\textit{\ (ii.b2)
Spacetime horns:}

If the base manifold $\Sigma_{4}$ has \emph{vanishing} Ricci scalar,%
\begin{equation}
\tilde{R}=0\ ,
\end{equation}
the solution is given by%
\begin{align}
f^{2}(r)  &  =\left(  a\sqrt{\sigma}r+\frac{1}{\sqrt{\sigma}r^{2}}\right)
^{2}\ ,\label{f-horn6}\\
g^{2}\left(  r\right)   &  =\sigma r^{2}\ , \label{grr-horn6}%
\end{align}
where $a$ is an integration constant.\textit{\newline}

\noindent$\cdot$\textit{\ (iii) Degeneracy:}

The base manifold $\Sigma_{4}$ is of constant curvature,
\begin{equation}
\tilde{R}_{\ kl}^{ij}=\gamma\delta_{kl}^{ij}\ ,
\end{equation}
and%
\begin{align}
g^{2}\left(  r\right)   &  =\sigma r^{2}+\gamma\ ,\\
f^{2}\left(  r\right)   &  :\text{\textrm{an arbitrary function.}}%
\end{align}

\bigskip

The purpose of this paper is extending this classification to higher
dimensions. The class of static metrics in Eq. (\ref{Ansatz}) with a base
manifold $\Sigma_{d-2}$, solves the EGB field equations $d>6$ dimensions
according to:

\subsection{$d\geq7$ dimensions}

\noindent$\mathbf{\circ}$\textbf{\ (i) Generic class:} For a generic value of
the Gauss-Bonnet coupling $a_{2}$, the most general solution of the EGB field
equations (\ref{egb}) within the class of metrics under consideration, given
by (\ref{Ansatz}), is such that:

The base manifold $\Sigma_{d-2}$ \emph{must be} Einstein,%
\begin{equation}
\tilde{R}_{~j}^{i}=\left(  d-3\right)  \gamma\delta_{j}^{i}\ ,\label{ccbm}%
\end{equation}
(with $\gamma$ normalized to $\pm1,0$), and simultaneously fulfills the
following (tensorial) condition on its Weyl tensor,%
\begin{equation}
\tilde{C}_{\ lm}^{ik}\tilde{C}_{\ jk}^{lm}=\frac{(d-3)!}{(d-6)!}(\xi
-\gamma^{2})\ \delta_{j}^{i}\ \text{,}\label{cond}%
\end{equation}
with{\small
\begin{equation}
f^{2}=g^{2}=\gamma+\frac{a_{1}}{a_{2}}r^{2}\left[  1\pm\sqrt{1-\frac
{a_{2}a_{0}}{a_{1}^{2}}+\frac{\mu}{r^{d-1}}+\frac{a_{2}^{2}}{a_{1}^{2}}%
\frac{\left(  \gamma^{2}-\xi\right)  }{r^{4}}}\right]  \label{fgeneric}%
\end{equation}
}where $\xi$ and $\mu$ are integration constants.\newline\noindent
$\mathbf{\circ}$\textbf{\ (ii) Special class:} If the Gauss-Bonnet coupling is
given by (\ref{sc}), there are three main branches of solutions in the bulk
according to the geometry of $\Sigma_{d-2}$:\newline

\noindent$\cdot$\textit{\ (ii.a.1) Black holes:}

The base manifold $\Sigma_{d-2}$ has the same restrictions as in the generic
case, i.e.,%
\begin{equation}
\tilde{R}_{~j}^{i}=\left(  d-3\right)  \gamma\delta_{j}^{i}\ ,
\end{equation}
and also fulfills%
\begin{equation}
\tilde{C}_{\ lm}^{ik}\tilde{C}_{\ jk}^{lm}=\frac{(d-3)!}{\left(  d-6\right)
!}\ (\xi-\gamma^{2})\ \delta_{j}^{i}\ \text{,}\label{condDG}%
\end{equation}
with $f^{2}$ and $g^{2}$ given by%
\begin{equation}
f^{2}=g^{2}=\gamma+\frac{a_{1}}{a_{2}}r^{2}\left[  1\pm\sqrt{\frac{\mu
}{r^{d-1}}+\frac{a_{2}^{2}}{a_{1}^{2}}\frac{\left(  \gamma^{2}-\xi\right)
}{r^{4}}}\right]  \label{HigherD-2a1}%
\end{equation}
where $\xi$ and $\mu$ are integration constants. Note that the asymptotic
behavior of (\ref{HigherD-2a1}) is slower than that of the generic case in
(\ref{fgeneric}).

\bigskip

For the remaining branches, the base manifold $\Sigma_{d-2}$ \emph{is no
longer restricted to be Einstein}, but instead fulfills a scalar condition:%
\begin{equation}
\tilde{H}+\frac{\gamma}{2}\frac{(d-4)!}{(d-7)!}\left[  \tilde{R}-\frac{\gamma
}{2}(d-2)(d-3)\right]  =0\ , \label{Trace-generic-D}%
\end{equation}
where $\tilde{H}$ is proportional to the Gauss-Bonnet invariant of
$\Sigma_{d-2}$, as defined in Eq. (\ref{s}), i.e.,%
\[
\tilde{H}:=\tilde{H}_{\ i}^{i}=\frac{\left(  6-d\right)  }{4}\left(  \tilde
{R}^{ijkl}\tilde{R}_{ijkl}-4\tilde{R}^{ij}\tilde{R}_{ij}+\tilde{R}^{2}\right)
\ ,
\]
and $g^{2}$ is given by%
\begin{equation}
g^{2}(r)=\sigma r^{2}+\gamma\hspace{1cm}\sigma:=\frac{a_{0}}{a_{1}}\ ,
\end{equation}
and $\gamma$ is a constant normalized to $\pm1,0$.\newline

\noindent$\cdot$\textit{\ (ii.a.2) Special black holes:}

The base manifold $\Sigma_{d-2}$, satisfies the Euclidean EGB equation for the
special case (\ref{sc}) in $d-2$ dimensions, i.e.,%
\begin{equation}
\tilde{H}_{\ j}^{i}-\gamma(d-5)(d-6)\tilde{G}_{\ j}^{i}-\frac{\gamma
^{2}(d-3)!}{4(d-7)!}\delta_{j}^{i}=0\ ,\label{ee}%
\end{equation}
admitting a unique maximally symmetric solution of curvature $\gamma$, whose
trace reduces to (\ref{Trace-generic-D}), and%
\[
f^{2}\left(  r\right)  =g^{2}\left(  r\right)  =\sigma r^{2}+\gamma\ .
\]
\newline

\noindent$\cdot$\textit{\ (ii.b1) Wormholes:}

The base manifold $\Sigma_{d-2}$ has constant nonvanishing Ricci scalar
\begin{equation}
\tilde{R}=(d-2)(d-3)\gamma, \label{ricciconstantarb}%
\end{equation}
and also satisfies the generic Euclidean EGB equation%
\begin{equation}
\tilde{H}_{\ j}^{i}-(d-5)(d-6)\left(  \gamma-\frac{J}{2}\right)  \tilde
{G}_{\ j}^{i}+\frac{\gamma\left(  J-\gamma\right)  }{4}\frac{(d-3)!}%
{(d-7)!}\delta_{j}^{i}=0\ , \label{wormarb1}%
\end{equation}
\bigskip where $J$ is an integration constant. Note that, by virtue of
(\ref{ricciconstantarb}) the trace of (\ref{wormarb1}) reduces to (\ref{ee})
giving no additional constraints on the geometry of $\Sigma_{d-2}$. The bulk
geometry is then determined by%
\[
g^{2}(r)=\sigma r^{2}+\gamma\ ,
\]
and $f^{2}\left(  r\right)  $ fulfills the generalized Legendre equation,
given by%
\begin{equation}
r\left(  \sigma r^{2}+\gamma\right)  f^{\prime\prime}+\left[  \left(
d-4\right)  \sigma r^{2}+\left(  d-5\right)  \gamma\right]  f^{\prime}-\left(
\left(  d-4\right)  \sigma r-\frac{\left(  d-5\right)  \left(  d-6\right)
J}{4r}\right)  f\left(  r\right)  =0\ . \label{ode}%
\end{equation}
The general solution then reads%
\begin{equation}
f^{2}\left(  r\right)  =r^{6-d}\left[  a\ P_{\nu}^{\ \mu}\left(
\sqrt{1+\gamma\sigma r^{2}}\right)  +b\ Q_{\nu}^{\ \mu}\left(  \sqrt
{1+\gamma\sigma r^{2}}\right)  \right]  ^{2} \label{Pll}%
\end{equation}
where $P_{\nu}^{\ \mu}\left(  x\right)  $ and $Q_{\nu}^{\ \mu}\left(
x\right)  $ are the generalized Legendre functions of first and second kind
respectively, with%
\begin{align}
\mu &  :=\frac{1}{2}\sqrt{\left(  d-6\right)  ^{2}-\frac{J}{\gamma}\left(
d-5\right)  \left(  d-6\right)  }\ ,\\
\nu &  :=\frac{d}{2}-2\ ,
\end{align}
and $a,$ $b$ are integration constants.\newline

\noindent$\cdot$\textit{\ (ii.b2) Spacetime horns:}

The base manifold $\Sigma_{d-2}$ has vanishing Ricci scalar%
\begin{equation}
\tilde{R}=0\ ,\label{ricciconstantarb0}%
\end{equation}
and also satisfies the Euclidean EGB equation devoid of the volume term%
\begin{equation}
\tilde{H}_{\ j}^{i}+\frac{J(d-5)(d-6)}{2}\tilde{G}_{\ j}^{i}%
=0\ ,\label{wormarb10}%
\end{equation}
\bigskip with $J$ an integration constant. As in the previous case, the
vanishing of the Ricci scalar of $\Sigma_{d-2}$, makes the trace of
(\ref{wormarb10}) reduce to (\ref{ee}) (with $\gamma=0$), without additional
conditions on $\Sigma_{d-2}$. The bulk geometry is given by%
\[
g^{2}(r)=\sigma r^{2}+\gamma\ ,
\]
and $f^{2}\left(  r\right)  $ fulfills Eq. (\ref{ode}) with $\gamma=0$, i.e.,%
\begin{equation}
\sigma r^{3}f^{\prime\prime}+\left(  d-4\right)  \sigma r^{2}f^{\prime
}-\left(  \left(  d-4\right)  \sigma r-\frac{\left(  d-5\right)  \left(
d-6\right)  J}{4r}\right)  f\left(  r\right)  =0\ ,
\end{equation}
whose general solution is%
\begin{equation}
f^{2}\left(  r\right)  :=r^{5-d}\left[  a\ J_{\alpha}\left(  \frac{1}{r}%
\sqrt{\frac{\left(  d-5\right)  \left(  d-6\right)  }{4\sigma}}\right)
+b\ Y_{\alpha}\left(  \frac{1}{r}\sqrt{\frac{\left(  d-5\right)  \left(
d-6\right)  }{4\sigma}}\right)  \right]  ^{2}\ .\label{Jll}%
\end{equation}
Here $J_{\alpha}\left(  x\right)  $ and $Y_{\alpha}\left(  x\right)  $ are the
Bessel functions of first and second kind respectively, with%
\begin{equation}
\alpha:=-\frac{d-3}{2}\ ,
\end{equation}
and $a,$ $b$ are integration constants.\newline

\noindent$\cdot$\textit{\ (iii) Degeneracy:}

The base manifold $\Sigma_{4}$ is of constant curvature,
\begin{equation}
\tilde{R}_{\ kl}^{ij}=\gamma\delta_{kl}^{ij}\ ,
\end{equation}
and%
\begin{align}
g^{2}\left(  r\right)   &  =\sigma r^{2}+\gamma\ ,\\
f^{2}\left(  r\right)   &  :\text{\textrm{an arbitrary function.}}%
\end{align}

\bigskip

This concludes the classification.

\section{Derivation of the classification scheme}

\label{derivation}

In order to proof the previous classification, it is convenient to work with
differential forms. The field equations for the EGB theory (\ref{Action EGB})
are given by (\ref{feq}), and in the case $a_{2}=0$ Eq. (\ref{feq}) reduces to
the Einstein equations with cosmological constant.

For the metric given in (\ref{Ansatz}) the vielbein can be chosen as
\[
e^{0}=f\left(  r\right)  dt\ ,\ e^{1}=\frac{dr}{g\left(  r\right)  }%
\ ,\ e^{m}=r\tilde{e}^{m}\ ,
\]
where $\tilde{e}^{m}$ is the vielbein of the base manifold $\Sigma_{d-2}$, so
that $m=2,3,...,d-1$, and the curvature two-form is then given by%
\begin{align}
R^{01}  &  =-\left(  gg^{\prime}\frac{f^{\prime}}{f}+g^{2}\frac{f^{\prime
\prime}}{f}\right)  e^{0}e^{1},\label{r1}\\
R^{0m}  &  =-\left(  g^{2}\frac{f^{\prime}}{fr}\right)  e^{0}e^{m},\\
R^{1m}  &  =-\frac{1}{2}\frac{\left(  g^{2}\right)  ^{\prime}}{r}e^{1}e^{m},\\
R^{mn}  &  =\tilde{R}^{mn}-\frac{g^{2}}{r^{2}}e^{m}e^{n}\ , \label{r4}%
\end{align}
where $\tilde{R}^{mn}$ stands for the curvature of $\Sigma_{d-2}$.

To proceed with the classification, we first solve the constraint
$\mathcal{E}_{0}=0$. One then finds that the analysis naturally splits in two
cases, one involving generic theories, and the other restricted to the special
class of theories defined by (\ref{sc}). Solving the remaining field equations
in each branch completes the classification.

\subsection{Solving the constraint}

The equation $\mathcal{E}_{0}=0$ reads
\begin{equation}
\epsilon_{m_{1}...m_{n}}\left[  a_{2}\left(  d-5\right)  \tilde{R}^{m_{1}%
m_{2}}\tilde{R}^{m_{3}m_{4}}+B_{0}\tilde{R}^{m_{1}m_{2}}\tilde{e}^{m_{3}%
}\tilde{e}^{m_{4}}+A_{0}\tilde{e}^{m_{1}}\tilde{e}^{m_{2}}\tilde{e}^{m_{3}%
}\tilde{e}^{m_{4}}\right]  \tilde{e}^{m_{5}}...\tilde{e}^{m_{n}}%
=0\ ,\label{eps0}%
\end{equation}
where $n=d-2$ is the dimension of the base manifold, and $A_{0}(r),\ B_{0}(r)$
are functions constructed out from$\ g^{2}(r)$ and its derivative (see
Appendix A). Taking a derivative of this equation with respect to $r$, one
obtains the following consistency condition%
\begin{equation}
\epsilon_{m_{1}...m_{n}}\left[  B_{0}^{\prime}\tilde{R}^{m_{1}m_{2}}%
+A_{0}^{\prime}\tilde{e}^{m_{1}}\tilde{e}^{m_{2}}\right]  \tilde{e}^{m_{3}%
}...\tilde{e}^{m_{n}}=0\ ,\label{baja2}%
\end{equation}
and since $\tilde{R}^{mn}$ and $\tilde{e}^{m}$ depend only on the coordinates
of $\Sigma_{n}$, one obtains that
\begin{equation}
A_{0}^{\prime}=-\gamma\;B_{0}^{\prime}\ ,\label{der}%
\end{equation}
where $\gamma$ is a constant. Eq.(\ref{der}) implies that%
\begin{equation}
A_{0}=-\gamma B_{0}-\left(  d-5\right)  a_{2}\xi\ ,\label{BAJ}%
\end{equation}
where $\xi$ is a new integration constant that has been conveniently rescaled.

Inserting (\ref{der}) in (\ref{baja2}) then gives the following condition%
\begin{equation}
B_{0}^{\prime}\;\epsilon_{m_{1}...m_{n}}\left(  \tilde{R}^{m_{1}m_{2}}%
-\gamma\tilde{e}^{m_{1}}\tilde{e}^{m_{2}}\right)  \tilde{e}^{m_{4}}%
...\tilde{e}^{m_{n}}=0\ , \label{difeps0}%
\end{equation}
which means that the analysis splits in two cases: $B_{0}^{\prime}\neq0$ and
$B_{0}^{\prime}=0$.

\subsubsection{The constraint $\mathcal{E}_{0}=0$ in the generic case
($B_{0}^{\prime}\neq0$)}

If $B_{0}^{\prime}$ is nonvanishing, the condition (\ref{difeps0}) reduces to%

\begin{equation}
\epsilon_{m_{1}...m_{n}}\left[  \tilde{R}^{m_{1}m_{2}}-\gamma\tilde{e}^{m_{1}%
}\tilde{e}^{m_{2}}\right]  \tilde{e}^{m_{3}}...\tilde{e}^{m_{n}}=0\ ,
\label{RicciScalarconstant}%
\end{equation}
which means that the Ricci scalar of the base manifold $\tilde{R}$ is a
constant, i.e.,%
\begin{equation}
\tilde{R}=n\left(  n-1\right)  \gamma\ . \label{rc}%
\end{equation}
Inserting (\ref{BAJ}) and (\ref{RicciScalarconstant}) in the constraint
(\ref{eps0}) gives and additional condition being quadratic in the curvature
of the base manifold:%
\begin{equation}
\left(  d-5\right)  a_{2}\;\epsilon_{m_{1}...m_{n}}\left(  \tilde{R}%
^{m_{1}m_{2}}\tilde{R}^{m_{3}m_{4}}-\xi\ \tilde{e}^{m_{1}}\tilde{e}^{m_{2}%
}\tilde{e}^{m_{3}}\tilde{e}^{m_{4}}\right)  \tilde{e}^{m_{5}}...\tilde
{e}^{m_{n}}=0\ . \label{gbm}%
\end{equation}
Equations (\ref{rc}) and (\ref{gbm}) restrict the geometry of $\Sigma_{n}$,
whereas (\ref{BAJ}) is a first order equation for $g^{2}\left(  r\right)  $
whose solution is
\begin{equation}
g^{2}\left(  r\right)  =\gamma+\frac{a_{1}}{a_{2}}r^{2}\left[  1\pm
\sqrt{1-\frac{a_{2}a_{0}}{a_{1}^{2}}+\frac{\mu}{r^{d-1}}+\frac{a_{2}^{2}%
}{a_{1}^{2}}\frac{\left(  \gamma^{2}-\xi\right)  }{r^{4}}}\right]  \ ,
\label{ggeneric}%
\end{equation}
with $\mu$ an integration constant.\newline

Note that we have \emph{not} assumed any relation between the coupling
constants of the theory, this is why these conditions apply in the
\emph{generic} case. \newline

\subsubsection{The constraint $\mathcal{E}_{0}=0$ in the special case
($B_{0}^{\prime}=0$)}

If $B_{0}^{\prime}$ vanishes Eq. (\ref{difeps0}) is trivially solved. On the
other hand, Eq. (\ref{der}) implies%
\begin{equation}
A_{0}^{\prime}=B_{0}^{\prime}=0\ ,
\end{equation}
and it is easy to see, from the expressions for $A_{0}$ and $B_{0}$ in the
Appendix, that this equation can be fulfilled only if the Gauss-Bonnet
coupling is fixed as%
\begin{equation}
a_{2}=\frac{a_{1}^{2}}{a_{0}}\ ,\label{tune}%
\end{equation}
which corresponds to the special class of theories (\ref{sc}). In this case
$g^{2}\left(  r\right)  $ is given by%
\begin{equation}
g^{2}\left(  r\right)  =\sigma r^{2}+\gamma\ ,\label{sg}%
\end{equation}
where we have defined
\begin{equation}
\sigma:=\frac{a_{0}}{a_{1}}\ .\label{sigma}%
\end{equation}
Therefore, since the functions $A_{0}$ and $B_{0}$ reduce to%
\begin{align}
A_{0} &  =\left(  d-5\right)  a_{2}\gamma^{2}\ ,\\
B_{0} &  =-2\left(  d-5\right)  a_{2}\gamma\ .
\end{align}
Eq. (\ref{eps0}) gives the following scalar restriction on the base manifold:%
\begin{equation}
\left(  d-5\right)  a_{2}\epsilon_{m_{1}...m_{n}}\left[  \tilde{R}^{m_{1}%
m_{2}}-\gamma\tilde{e}^{m_{1}}\tilde{e}^{m_{2}}\right]  \left[  \tilde
{R}^{m_{3}m_{4}}-\gamma\tilde{e}^{m_{3}}\tilde{e}^{m_{4}}\right]  \tilde
{e}^{m_{5}}...\tilde{e}^{m_{n}}=0\ .\label{trEGBtuneado}%
\end{equation}
Note that this last condition on $\Sigma_{n}$ is weaker than the ones obtained
in the generic case (\ref{rc}) and (\ref{gbm}). One should keep in mind that
Eq. (\ref{trEGBtuneado}) applies only for the \emph{special theories
}fulfilling (\ref{tune}).

\subsection{Solving the remaining equations}

The equation $\mathcal{E}_{1}=0$ reduces to
\begin{equation}
\epsilon_{m_{1}...m_{n}}\left[  \left(  d-5\right)  a_{2}\tilde{R}^{m_{1}%
m_{2}}\tilde{R}^{m_{3}m_{4}}+B_{1}\ \tilde{R}^{m_{1}m_{2}}\tilde{e}^{m_{3}%
}\tilde{e}^{m_{4}}+A_{1}\ \tilde{e}^{m_{1}}\tilde{e}^{m_{2}}\tilde{e}^{m_{3}%
}\tilde{e}^{m_{4}}\right]  \tilde{e}^{m_{5}}...\tilde{e}^{m_{n}}=0\ ,
\label{eps1}%
\end{equation}
where $A_{1}$ and $B_{1}$ are functions of $r$, $f$, $g$ and their derivatives
(see Appendix A). Subtracting (\ref{eps1}) from (\ref{eps0}), the quadratic
terms cancel out, and we obtain
\begin{equation}
\epsilon_{m_{1}...m_{n}}\left[  \left(  B_{0}-B_{1}\right)  \ \tilde{R}%
^{m_{1}m_{2}}+\left(  A_{0}-A_{1}\right)  \ \tilde{e}^{m_{1}}\tilde{e}^{m_{2}%
}\right]  \tilde{e}^{m_{3}}...\tilde{e}^{m_{n}}=0\ . \label{e0-e1}%
\end{equation}
The projection of the EGB field equations (\ref{feq}) on $\Sigma_{n}$,
$\mathcal{E}_{m}=0$, reads%
\begin{equation}
\epsilon_{mm_{2}...m_{n}}\left[  \left(  d-5\right)  \left(  d-6\right)
a_{2}\tilde{R}^{m_{2}m_{3}}\tilde{R}^{m_{4}m_{5}}+C\ \tilde{R}^{m_{2}m_{3}%
}\tilde{e}^{m_{4}}\tilde{e}^{m_{5}}+D\ \tilde{e}^{m_{2}}\tilde{e}^{m_{3}%
}\tilde{e}^{m_{4}}\tilde{e}^{m_{5}}\right]  \tilde{e}^{m_{6}}...\tilde
{e}^{m_{n}}=0\ . \label{epsm}%
\end{equation}
Where again $C$ and $D$ are functions of $r$, $f$, $g$ and their derivatives,
given in Appendix A.\newline

We will solve (\ref{e0-e1}) and (\ref{epsm}) for the generic and special cases separately.

\subsubsection{Radial and angular equations: Generic case}

Introducing (\ref{RicciScalarconstant}) in (\ref{e0-e1}) we obtain%
\begin{equation}
\left(  B_{0}-B_{1}\right)  \ \gamma+\left(  A_{0}-A_{1}\right)  =0\ ,
\end{equation}
which reduces to%
\begin{equation}
\frac{d}{dr}\left[  \ln\frac{g\left(  r\right)  }{f\left(  r\right)  }\right]
\left[  g^{2}-\left(  \sigma r^{2}+\gamma\right)  \right]  =0\ ,
\label{eps1generic}%
\end{equation}

Note that since in the generic case the function $g^{2}$ is given by
(\ref{ggeneric}), the second factor in (\ref{eps1generic}) does not vanish in
general. This implies that $f^{2}\left(  r\right)  $ is proportional to
$g^{2}\left(  r\right)  $, and the constant of proportionality can be
reabsorbed by a time rescaling, so that%
\begin{equation}
f^{2}\left(  r\right)  =g^{2}\left(  r\right)  \ , \label{f=g}%
\end{equation}
where $g^{2}(r)$ is given in (\ref{ggeneric}).

\bigskip

Let us now solve the remaining equations $\mathcal{E}_{m}=0$. By virtue of
(\ref{f=g}), the functions $C$ and $D$ fulfill the following relation%
\begin{equation}
D=-\gamma C-\left(  d-5\right)  \left(  d-6\right)  a_{2}\xi\ . \label{DC}%
\end{equation}

Taking a derivative of (\ref{epsm}) with respect to $r$ we obtain
\begin{equation}
C^{\prime}\epsilon_{mm_{2}...m_{n}}\left[  \tilde{R}^{m_{2}m_{3}}%
-\gamma\ \tilde{e}^{m_{2}}\tilde{e}^{m_{3}}\right]  \tilde{e}^{m_{4}}%
...\tilde{e}^{m_{n}}=0\ , \label{drepsm}%
\end{equation}
and since it is straightforward to check that $C^{\prime}\neq0$ for the
generic case, this equation is solved provided%
\begin{equation}
\epsilon_{mm_{2}...m_{n}}\left[  \tilde{R}^{m_{2}m_{3}}-\gamma\ \tilde
{e}^{m_{2}}\tilde{e}^{m_{3}}\right]  \tilde{e}^{m_{4}}...\tilde{e}^{m_{n}%
}=0\ , \label{cond1}%
\end{equation}
which means that the base manifold must be Einstein.

Furthermore, if we use the latter equation and (\ref{DC}), then Eq.
(\ref{epsm}) reads%
\begin{equation}
\left(  d-5\right)  \left(  d-6\right)  a_{2}\epsilon_{mm_{2}...m_{n}}\left[
\tilde{R}^{m_{2}m_{3}}\tilde{R}^{m_{4}m_{5}}-\xi\ \tilde{e}^{m_{2}}\tilde
{e}^{m_{3}}\tilde{e}^{m_{4}}\tilde{e}^{m_{5}}\right]  \tilde{e}^{m_{6}%
}...\tilde{e}^{m_{n}}=0\ . \label{cond2}%
\end{equation}

It is simple to verify (see Appendix B) that for and Einstein manifold
(\ref{cond1}), this last equation reduces to (\ref{condDG}).

\bigskip

This concludes the proof of the classification in the generic case
\textbf{(i)}, which includes the case \textbf{(ii.a.1)} when the condition
(\ref{sc}) is further fulfilled.

\subsubsection{Radial and angular equations: Special case}

Using (\ref{sg}) in (\ref{e0-e1}) gives
\begin{equation}
a_{2}\frac{d}{dr}\left[  \ln\frac{g\left(  r\right)  }{f\left(  r\right)
}\right]  \ \epsilon_{m_{1}...m_{n}}\left[  \ \tilde{R}^{m_{1}m_{2}}%
-\gamma\ \tilde{e}^{m_{1}}\tilde{e}^{m_{2}}\right]  \tilde{e}^{m_{3}}%
...\tilde{e}^{m_{n}}=0\ , \label{e0-e1special}%
\end{equation}
which means that the analysis splits in the following two cases:\newline

\noindent\textbf{(ii.a.2): }This is the case where the first factor in
(\ref{e0-e1special}) vanishes. Hence, after a rescaling of time, one obtains%
\begin{equation}
f^{2}\left(  r\right)  =g^{2}\left(  r\right)  =\sigma r^{2}+\gamma\ ,
\label{f=g=spec}%
\end{equation}
with $\sigma$ given by (\ref{sigma}). Replacing (\ref{f=g=spec}) in
$\mathcal{E}_{m}=0$, implies that the metric of the base manifold fulfills the
following equation%
\begin{equation}
\left(  d-5\right)  \left(  d-6\right)  a_{2}\epsilon_{mm_{2}...m_{n}}\left[
\tilde{R}^{m_{2}m_{3}}-\gamma\tilde{e}^{m_{2}}\tilde{e}^{m_{3}}\right]
\left[  \tilde{R}^{m_{4}m_{5}}-\gamma\tilde{e}^{m_{4}}\tilde{e}^{m_{5}%
}\right]  \tilde{e}^{m_{6}}...\tilde{e}^{m_{n}}=0\ . \label{weakest}%
\end{equation}
It is worth pointing out that Eq.\ (\ref{weakest}) is the same (Euclidean) EGB
equation for the special case (\ref{tune}), but in $n=d-2$ dimensions. Once
expressed in terms of tensors in $d>6$ dimensions, Eq. (\ref{weakest}) reads%
\begin{equation}
\delta_{jl_{1}l_{2}l_{3}l_{4}}^{ik_{1}k_{2}k_{3}k_{4}}\left(  \tilde
{R}_{\ k_{1}k_{2}}^{l_{1}l_{2}}-\gamma\delta_{k_{1}k_{2}}^{l_{1}l_{2}}\right)
\left(  \tilde{R}_{\ k_{3}k_{4}}^{l_{3}l_{4}}-\gamma\delta_{k_{3}k_{4}}%
^{l_{3}l_{4}}\right)  =0\ , \label{deltaRR}%
\end{equation}
which reduces to (\ref{ee}). This corresponds to the case \textbf{(ii.a.2)} of
the classification.\newline

\noindent\textbf{(ii.b) and (iii): }In the case when the first factor of
(\ref{e0-e1special}) does not vanish, i.e., when $f\left(  r\right)  $ is
\emph{not} proportional to $g\left(  r\right)  $, equation (\ref{e0-e1special}%
) reduces to%
\begin{equation}
\epsilon_{m_{1}...m_{n}}\left[  \ \tilde{R}^{m_{1}m_{2}}-\gamma\ \tilde
{e}^{m_{1}}\tilde{e}^{m_{2}}\right]  \tilde{e}^{m_{3}}...\tilde{e}^{m_{n}%
}=0\ ,
\end{equation}
which means that the Ricci scalar of $\Sigma_{n}$ is a constant,
\begin{equation}
\tilde{R}=n\left(  n-1\right)  \gamma\ .
\end{equation}

The \textquotedblleft angular\textquotedblright\ equation (\ref{epsm}) in this
case reads%
\begin{gather}
\mathcal{E}_{m}:=\epsilon_{mm_{2}...m_{n}}\left[  \tilde{R}^{m_{2}m_{3}%
}-\gamma\tilde{e}^{m_{2}}\tilde{e}^{m_{3}}\right]  \left[  \tilde{R}%
^{m_{4}m_{5}}-\gamma\tilde{e}^{m_{4}}\tilde{e}^{m_{5}}\right]  \tilde
{e}^{m_{6}}...\tilde{e}^{m_{n}}\nonumber\\
+\frac{\mathcal{D}\left[  f\left(  r\right)  \right]  }{f(r)}\ \epsilon
_{mm_{2}...m_{n}}\left[  \tilde{R}^{m_{2}m_{3}}-\gamma\tilde{e}^{m_{2}}%
\tilde{e}^{m_{3}}\right]  \tilde{e}^{m_{4}}...\tilde{e}^{m_{n}}=0\ ,
\label{epsmspecial}%
\end{gather}
where $\mathcal{D}$ is the linear differential operator defined by%
\begin{equation}
\mathcal{D}\left[  f\left(  r\right)  \right]  :=\frac{4}{\left(  d-5\right)
\left(  d-6\right)  }\left[  -r^{2}\left(  \sigma r^{2}+\gamma\right)
f^{\prime\prime}-r\left(  \left(  d-4\right)  \sigma r^{2}+\left(  d-5\right)
\gamma\right)  f^{\prime}+\sigma r^{2}\left(  d-4\right)  f\left(  r\right)
\right]  \ . \label{dop}%
\end{equation}

\bigskip

Taking a derivative of equation (\ref{epsmspecial}) with respect to $r$ leads
us to consider the following two subcases:\newline

\noindent\textbf{(ii.b): }If $\mathcal{D}\left[  f\left(  r\right)  \right]
=Jf(r)$, where $J$ is a constant, then (\ref{epsmspecial}) reduces to%
\begin{align}
&  \epsilon_{mm_{2}...m_{n}}\left[  \tilde{R}^{m_{2}m_{3}}-\gamma\tilde
{e}^{m_{2}}\tilde{e}^{m_{3}}\right]  \left[  \tilde{R}^{m_{4}m_{5}}%
-\gamma\tilde{e}^{m_{4}}\tilde{e}^{m_{5}}\right]  \tilde{e}^{m_{6}}%
...\tilde{e}^{m_{d-3}}\nonumber\\
&  +J\ \epsilon_{mm_{2}...m_{n}}\left[  \tilde{R}^{m_{2}m_{3}}-\gamma\tilde
{e}^{m_{2}}\tilde{e}^{m_{3}}\right]  \tilde{e}^{m_{4}}...\tilde{e}^{m_{d-3}%
}=0\ .\label{epsmespecilnodiff}%
\end{align}
This means that the base manifold also fulfills an Euclidean EGB equation for
a generic choice of the Gauss-Bonnet coupling in $n=d-2$ dimensions, where the
constant $J$ measures the departure of (\ref{epsmespecilnodiff}) from the
special case. The function $f\left(  r\right)  $ solves the following
equation:%
\begin{equation}
r^{2}\left(  \sigma r^{2}+\gamma\right)  f^{\prime\prime}+r\left(  \left(
d-4\right)  \sigma r^{2}+\left(  d-5\right)  \gamma\right)  f^{\prime}+\left(
\frac{\left(  d-5\right)  \left(  d-6\right)  J}{4}-\sigma\left(  d-4\right)
r^{2}\right)  f=0\label{ecfworm}%
\end{equation}
whose integration depends on the value of $\gamma$.

$\bullet$ \textbf{(ii.b.1): }For $\gamma\neq0$ the solution of (\ref{ecfworm})
is given by%
\[
f\left(  r\right)  =r^{3-\frac{d}{2}}\left[  a\ P_{\nu}^{\ \mu}\left(
\sqrt{\gamma\sigma r^{2}+1}\right)  +b\ Q_{\nu}^{\ \mu}\left(  \sqrt
{\gamma\sigma r^{2}+1}\right)  \right]
\]
where $P_{\nu}^{\ \mu}\left(  x\right)  $ and $Q_{\nu}^{\ \mu}\left(
x\right)  $ are the generalized Legendre functions of first and second kind
respectively, with%
\begin{align}
\mu &  :=\frac{1}{2}\sqrt{\left(  d-6\right)  ^{2}-\frac{J}{\gamma}\left(
d-5\right)  \left(  d-6\right)  }\ ,\\
\nu &  :=\frac{d}{2}-2\ ,
\end{align}
and $a,$ $b$ are integration constants.

$\bullet$ \textbf{(ii.b.2):} For $\gamma=0$, equation (\ref{ecfworm})
integrates as%
\begin{equation}
f\left(  r\right)  :=r^{-\frac{d-5}{2}}\left[  a\ J_{\alpha}\left(  \frac
{1}{r}\sqrt{\frac{\left(  d-5\right)  \left(  d-6\right)  }{4\sigma}}\right)
+b\ Y_{\alpha}\left(  \frac{1}{r}\sqrt{\frac{\left(  d-5\right)  \left(
d-6\right)  }{4\sigma}}\right)  \right]  \ ,
\end{equation}
where $J_{\alpha}\left(  x\right)  $ and $Y_{\alpha}\left(  x\right)  $ are
the Bessel functions of first and second kind respectively, with%
\begin{equation}
\alpha:=-\frac{d-3}{2}\ .
\end{equation}
This concludes the proof corresponding to the cases \textbf{(ii.b.1)} and
\textbf{(ii.b.2)}.

\noindent\textbf{(iii) }If $\mathcal{D}\left[  f\left(  r\right)  \right]
/f(r)$ is not a constant, then Eq. (\ref{epsmspecial}) is solved provided the
base manifold simultaneously fulfills the Einstein and the EGB equations in
the special case with the same cosmological constant, i.e.,%
\begin{equation}
\epsilon_{mm_{2}...m_{n}}\left[  \tilde{R}^{m_{2}m_{3}}-\gamma\tilde{e}%
^{m_{2}}\tilde{e}^{m_{3}}\right]  \tilde{e}^{m_{4}}...\tilde{e}^{m_{n}}=0\ ,
\label{cc1}%
\end{equation}%
\begin{equation}
\epsilon_{mm_{2}...m_{n}}\left[  \tilde{R}^{m_{2}m_{3}}-\gamma\tilde{e}%
^{m_{2}}\tilde{e}^{m_{3}}\right]  \left[  \tilde{R}^{m_{4}m_{5}}-\gamma
\tilde{e}^{m_{4}}\tilde{e}^{m_{5}}\right]  \tilde{e}^{m_{6}}...\tilde
{e}^{m_{n}}=0\ ; \label{cc2}%
\end{equation}
and $f\left(  r\right)  $ becomes an arbitrary function.\newline For an
Euclidean Einstein manifold fulfilling (\ref{cc1}), Eq. (\ref{cc2}) reduces to%
\begin{equation}
\tilde{C}_{\ lm}^{ij}\tilde{C}_{\ jk}^{lm}=0\ ,
\end{equation}
which implies that $\Sigma_{n}$ must be of constant curvature $\gamma$ (see
Appendix B), i.e.,%
\begin{equation}
\tilde{R}_{\ kl}^{ij}=\gamma\delta_{kl}^{ij}\ .
\end{equation}

\bigskip

This ends the proof of the classification.

\section{Discussion}

In this paper, the class of static metrics given by (\ref{Ansatz}) that solves
the EGB field equations in vacuum for $d\geq5$ dimensions has been classified.
It was shown that for a generic value of the Gauss-Bonnet coupling, the base
manifold must be necessarily Einstein, with an additional restriction on its
Weyl tensor if $d>5$. The boundary admits a wider class of geometries only in
the special case when the Gauss-Bonnet coupling is given by (\ref{sc}), such
that the theory admits a unique maximally symmetric solution. The additional
freedom in the boundary metric enlarges the class of allowed geometries in the
bulk, which are classified within three main branches, containing new black
holes and wormholes in vacuum.

\bigskip

In the five-dimensional case, the classification was performed in
\cite{DOT5d}, including a thorough analysis of the geometrically well-behaved
solutions including black holes, wormholes and spacetime horns. It was also
shown that these solutions have finite Euclidean action (regularized through
the boundary terms proposed in \cite{MOTZ1, MOTZ2}), which reduces to the free
energy in the case of black holes, and vanishes in the remaining cases. The
mass was also obtained from the corresponding conserved charge written as a
surface integral. For a generic choice of the Gauss-Bonnet coupling, the
solution was obtained in \cite{CaienAdS} assuming the base manifold to be of
constant curvature, and in the spherically symmetric case Eq.
(\ref{gcuadradogenerico}) reduces to the well-known solution of Boulware and
Deser \cite{BD}. In the special case, in which the Gauss-Bonnet coupling is
given by (\ref{sc}), the Lagrangian can be written as a Chern-Simons form
\cite{Chamseddine} and its locally supersymmetric extension is known
\cite{CHAM2,sugraricalld}. For the special case, when the cosmological
constant is negative $\left(  \sigma>0\right)  $ this solution, corresponding
to the branch (ii.a), describes a black hole \cite{Cai-Soh}, \cite{ATZ}, which
for spherical symmetry, reduces to the one found in \cite{BD}, \cite{BTZ}. It
can also be seen that the black hole metric still solves the field equations
even in the presence of a nontrivial fully antisymmetric torsion
\cite{Camello-Fabrizio-Troncoso}. For the branch (ii.b.1) with $\gamma=-1$,
the solution with $|a|<1$ corresponds to the wormhole in vacuum found in
\cite{DOTWorm}. It has also been shown that if the base manifold is given by
the hyperbolic space in three dimensions, i.e., $\Sigma_{3}=H_{3}$ with no
identifications, this metric describes a smooth gravitational soliton
\cite{COT}. If $|a|=1$, the solution reduces to a different kind of wormholes
possessing inequivalent asymptotic regions. For the branch (ii.b.2), if
$a\geq0$ the solution describes a \textquotedblleft spacetime
horn\textquotedblright\ \cite{DOT5d}.

\bigskip

In the six-dimensional case, the classification was carried out in \cite{JP}.
For a generic choice of the Gauss-Bonnet coupling, besides the mass parameter
$\mu$, an independent integration constant $\xi$ appears. The base manifold
$\Sigma_{4}$ has to be Einstein with an additional scalar condition on its
geometry, given by (\ref{E4propxi}), which means that the Euler density of
$\Sigma_{4}$ must be constant. Therefore, if one assumes that $\Sigma_{4}$ is
compact and without boundary, integration of Eq.(\ref{E4propxi}) on
$\Sigma_{4}$ gives a topological restriction on the base manifold,
constraining the new parameter to be $\xi=\frac{4}{3}\pi^{2}\frac{\chi
(\Sigma_{4})}{\mathcal{V}_{4}}$, where $\chi(\Sigma_{4})$ is the Euler
characteristic of the base manifold and $\mathcal{V}_{4}$ stands for its
volume. Note that the term proportional to $r^{-4}$ inside the square root in
the metric (\ref{generic6}) vanishes if and only if the base manifold is of
constant curvature. It is worth pointing out that this term severely modifies
the asymptotic behavior of the metric. Depending on the value of the
parameters, this spacetime can describe black holes being asymptotically
locally (A)dS or flat. The asymptotic behaviour of the metric is further
relaxed in the special case (\ref{sc}) (see (ii.a.1)), which for a constant
curvature base manifold $\Sigma_{4}$ reduces to the solution found in
\cite{ATZ}.

When (\ref{sc}) is fulfilled, it was shown that the restriction that
$\Sigma_{4}$ be Einstein can be circumvented (case (ii)). For the case
(ii.a.2), the geometry of the base manifold is as relaxed as possible, since
it has to fulfill just a single scalar equation, given by (\ref{paratodas6}).
Remarkably, if the Ricci scalar is further required to be a nonvanishing
constant $\gamma=\pm1$, for negative cosmological constant, wormholes in
vacuum also exist in six dimensions, provided $a^{2}<\frac{\pi^{2}}{4}$ (case
(ii.b.1) with $\gamma=-1$), and the volume of the base manifold turns out to
be fixed in terms of the Euler characteristic, according to $\chi(\Sigma
_{4})=\frac{3}{4\pi^{2}}\mathcal{V}_{4}$. In the case of $\gamma=0$, i.e., if
the base manifold $\Sigma_{4}$ has vanishing Ricci scalar, one obtains that
$\chi(\Sigma_{4})=0$, and for $a\geq0$ the metric looks like a
\textquotedblleft spacetime horn\textquotedblright. In the six-dimensional
case this classification has been further explored in \cite{CZ} for the case
in which the functions $f^{2}$ and $g^{2}$, are also time-dependent.

\bigskip

The classification of these solutions presents special features in $d=5$ and
$6$ dimensions, and as explained above, a common pattern arises in higher
dimensions. In the case of $d\geq7$, for a generic choice of the Gauss-Bonnet
coupling (case (i)) it was found that the base manifold has to be Einstein,
fulfilling the additional condition (\ref{cond}), in agreement with
\cite{Dotti-Gleiser}. Apart from the mass parameter $\mu$, an additional
integration constant $\xi$ appears. For spherical symmetry, one recovers the
result found by Boulware and Deser \cite{BD}. The gravitational stability in
the spherically symmetric case was analyzed in \cite{stab}, where it was found
that, contrary to what happens for higher-dimensional spherical black holes in
GR, in the asymptotically flat five and six-dimensional cases, there is a
critical mass below which the black holes become unstable \cite{stab2,stab3}.
If the base manifold $\Sigma_{d-2}$ is of constant curvature, then the
condition (\ref{cond}) implies that $\gamma^{2}-\xi=0$, and one recovers the
results found by Cai \cite{CaienAdS}. The difference $\gamma^{2}-\xi$,
parametrizes the deviation of the base manifold from being of constant
curvature, and it is worth pointing out that if $\xi\neq\gamma^{2}$, the
metric given by (\ref{fgeneric}) acquires an additional term of order $r^{-4}$
within the square root in (\ref{fgeneric}) regardless the spacetime dimension,
so that the metric possesses a slower fall off at infinity as compared with
ones with base manifolds of constant curvature. In the asymptotic region, the
behaviour of the metric is further relaxed in the special case (\ref{sc})
\cite{BHS}, see (ii.a.2), and for base manifolds $\Sigma_{d-2}$ of constant
curvature, the solution reduces to the one found in \cite{ATZ}.

It was shown that for the special choice (\ref{sc}), the restriction that
$\Sigma_{d-2}$ be Einstein can be surmounted (case (ii)). In the case
(ii.a.2), the geometry of the base manifold turns out to be as relaxed as
possible, since it just has to fulfill just Eq. (\ref{ee}), corresponding to
the Euclidean EGB equation for the special case (\ref{sc}) in $d-2$
dimensions, admitting a unique maximally symmetric solution of curvature
$\gamma$.

For the choice (\ref{sc}), if the base manifold $\Sigma_{d-2}$ has constant
Ricci scalar and fulfills the Euclidean EGB equation in $d-2$ dimensions in
(\ref{wormarb1}), which depends on an integration constant $J$, one recovers
cases (ii.b) for which the metric is expressed in terms of generalized
Legendre functions for $\gamma=\pm1$ (Eq. (\ref{Pll}) of case (ii.b.1)), and
Bessel functions for $\gamma=0$ (Eq. (\ref{Jll}) of case (ii.b.1)). In the
case of $J=0$ Eq. (\ref{wormarb1}) reduces to (\ref{ee}), and the metric
explicitly acquires the following form%
\begin{equation}
f\left(  r\right)  =ar+\frac{b}{r^{d-4}}\ ,
\end{equation}
for $\gamma=0$, and%
\begin{equation}
f\left(  r\right)  =a\sqrt{\sigma r^{2}+\gamma}+b\ \frac{h\left(  r\right)
}{r^{d-6}}\ ,
\end{equation}
with%
\begin{equation}
h\left(  r\right)  =\left\{
\begin{array}
[c]{ccc}%
2\sigma r^{2}+\gamma & : & d=7\\
3\sigma r^{2}+\gamma-3\sigma r^{2}\sqrt{\gamma\sigma r^{2}+1}\tanh^{-1}\left(
\left(  \gamma\sigma r^{2}+1\right)  ^{-1/2}\right)  & : & d=8\\
8\sigma^{2}r^{4}+4\sigma\gamma r^{2}-1 & : & d=9\\
2-5\sigma r^{2}\gamma-15\sigma^{2}r^{4}+15\sigma r^{4}\sqrt{\gamma\sigma
r^{2}+1}\tanh^{-1}\left(  \left(  \gamma\sigma r^{2}+1\right)  ^{-1/2}\right)
& : & d=10\\
16\sigma^{3}r^{6}+8\sigma^{2}\gamma r^{4}-2\sigma\gamma^{2}r^{2}-\gamma^{3} &
: & d=11
\end{array}
\right.  \ .
\end{equation}
for $\gamma=\pm1$.

\bigskip

As explained in \cite{DOT5d}, \cite{JP}, in five and six dimensions
respectively, and extended here to any dimension $d\geq7$, for the EGB\ theory
with special choice of the Gauss-Bonnet coupling (\ref{sc}), metrics of the
form
\begin{equation}
ds^{2}=-f^{2}\left(  r\right)  dt^{2}+\frac{dr^{2}}{\sigma r^{2}+\gamma}%
+r^{2}d\Sigma_{(d-2)}^{2}\ ,\label{Deg}%
\end{equation}
with $\sigma=\frac{a_{0}}{a_{1}}$, and $\gamma=\pm1,0$, may acquire degeneracy
(case (iii)). Degeneracy occurs for the metric (\ref{Deg}) when the base
manifold $\Sigma_{d-2}$ is of constant curvature $\gamma$, since the EGB
equations turn out to be solved \emph{for an arbitrary function} $f^{2}\left(
r\right)  $. Thus, in particular, the Lifshitz spacetimes in \cite{MS},
\cite{Pang}, \cite{DM}, fall within this class.

This kind of degeneracy is a known feature of a wide class of theories
\cite{dege}. A similar degeneracy has been found in the context of Birkhoff's
theorem for the EGB theory in vacuum \cite{Zegers}, \cite{Deser}, and also for
theories containing dilaton and an axion fields coupled with a Gauss-Bonnet
term \cite{ACD}.

\bigskip

From the point of view of the AdS/CFT correspondence \cite{MAGOO}, the dual
CFT is expected to have a behaviour that strongly dependends on the choice of
the base manifold $\Sigma_{d-2}$. Note that the existence of wormholes with
AdS asymptotics, as the ones reported here, raises some puzzles within this
context \cite{WY}, \cite{MM}, \cite{AOP}. Nevertheless, in five dimensions,
some interesting results have been found in \cite{stringsonAdSWorms}. The EGB
theory also admits wormhole solutions in the presence of matter that fulfill
the standard energy conditions \cite{Bhawal-Kar, HIDEKI2} \cite{Simeone1,
Simeone2}. From the gravity side of the correspondence, the addition of a
Gauss-Bonnet term in the action has recently attracted a lot of attention
concerning the hydrodynamic limit of the dual CFT \cite{BU1}-\cite{CAINHE}.

\bigskip

The EGB theory in also possesses rotating solutions with a nontrivial geometry
at the boundary \cite{NATH}. Currently, a wide spectrum of solutions in vacuum
is known, including black strings and black p-branes \cite{GOT,KM,Terrence},
spontaneous compactifications \cite{MH}-\cite{CaiCOMP}, metrics with a
nontrivial jump in the extrinsic curvature \cite{HTronZ,GW,GGGW}, and even
solutions with nontrivial torsion
\cite{ArosContreras,Camello1,Camello-Fabrizio-Troncoso}.

\section{Acknowledgements}

We thank Steve Willison for helpful comments. This research is partially
funded by Fondecyt grants N%
${{}^o}$
1085322, 1095098, 11090281, by the Conicyt grant \textquotedblleft
Southern Theoretical Physics Laboratory\textquotedblright\ ACT-91,
by the CONICET grant PIP 112-200801-02479 from CONICET and by 
grants number  05/B384 and 05/B253 from Universidad Nacional de C\'ordoba. 
GD is supported by CONICET. 
The Centro
de Estudios Cient\'{\i}ficos (CECS) is funded by the Chilean Government
through the Millennium Science Initiative and the Centers of Excellence Base
Financing Program of Conicyt. CECS is also supported by a group of private
companies which at present includes Antofagasta Minerals, Arauco, Empresas
CMPC, Indura, Naviera Ultragas and Telef\'{o}nica del Sur. CIN is funded by
Conicyt and the Gobierno Regional de Los R\'{\i}os.
\appendix

\section{Functions appearing in the field equations}

Here we present the expressions for the functions appearing in the EGB field
equations. For the constraint $\mathcal{E}_{0}=0$ in (\ref{eps0}), the
corresponding functions are defined by
\begin{align}
A_{0}  &  :=r^{-d+6}\left[  a_{0}r^{d-1}-2a_{1}r^{d-3}g^{2}+a_{2}r^{d-5}%
g^{4}\right]  ^{\prime}\ ,\\
B_{0}  &  :=2r^{-d+6}\left[  a_{1}r^{d-3}-a_{2}r^{d-5}g^{2}\right]  ^{\prime
}\ ,
\end{align}
and for the radial equation $\mathcal{E}_{1}=0$ (\ref{eps1}) those are
\begin{align}
A_{1}\left(  r\right)   &  :=a_{0}\left(  d-1\right)  r^{4}-2a_{1}g^{2}%
r^{2}\left(  \left(  d-3\right)  +2\frac{f^{\prime}}{f}r\right) \nonumber\\
&  +a_{2}g^{4}\left(  \left(  d-5\right)  +4\frac{f^{\prime}}{f}r\right)
\ ,\label{a1}\\
B_{1}\left(  r\right)   &  :=-2a_{2}g^{2}\left(  \left(  d-5\right)
+2\frac{f^{\prime}}{f}r\right)  +2\left(  d-3\right)  a_{1}r^{2}\ . \label{b1}%
\end{align}

For the projection of the EGB field equations along the base manifold
$\Sigma_{n}$, $\mathcal{E}_{m}=0$ in (\ref{epsm}), the corresponding functions
are given by%
\begin{align}
C  &  :=-2a_{2}r^{2}\left[  \left(  g^{2}\right)  ^{\prime}\frac{f^{\prime}%
}{f}+2g^{2}\frac{f^{\prime\prime}}{f}+\left(  d-5\right)  \left(  2g^{2}%
\frac{f^{\prime}}{rf}+r^{5-d}\left(  g^{2}r^{d-6}\right)  ^{\prime}\right)
\right] \nonumber\\
&  +2\left(  d-3\right)  \left(  d-4\right)  a_{1}r^{2}\ ,\label{c}\\
D  &  :=\left(  d-1\right)  \left(  d-2\right)  a_{0}r^{4}-2a_{1}r^{2}\left[
\left(  d-3\right)  r^{5-d}\left(  g^{2}r^{d-4}\right)  ^{\prime}\right.
\nonumber\\
&  \left.  +\frac{r}{f}\left(  2\left(  d-3\right)  g^{2}f^{\prime}+\left(
g^{2}\right)  ^{\prime}f^{\prime}r+2g^{2}f^{\prime\prime}r\right)  \right]
+a_{2}r\left[  \left(  d-5\right)  r^{6-d}\left(  g^{4}r^{d-6}\right)
^{\prime}\right. \nonumber\\
&  \left.  +4\left(  d-5\right)  g^{4}\frac{f^{\prime}}{f}+2\left(
g^{4}\right)  ^{\prime}\frac{f^{\prime}}{f}r+\left(  g^{4}\right)  ^{\prime
}\frac{f^{\prime}}{f}r+4g^{4}\frac{f^{\prime\prime}}{f}r\right]  \ . \label{d}%
\end{align}

\section{Some useful geometrical identities}

An $n-$dimensional Einstein manifold $\Sigma_{n}$, fulfills%
\begin{equation}
R_{\ j}^{i}=\gamma(n-1)\;\delta_{j}^{i}\ . \label{em}%
\end{equation}
In this case the Einstein tensor reads%
\begin{equation}
G_{j}^{i}=-\frac{(n-2)(n-1)}{2}\gamma\;\delta_{j}^{i}\ . \label{eet}%
\end{equation}
and for $n>3$ the Weyl tensor defined as the \textquotedblleft trace free
part\textquotedblright\ of the Riemann tensor%
\begin{equation}
C_{\ kl}^{ij}:=R_{\ kl}^{ij}-\frac{4}{n-2}\delta_{\lbrack k}^{[i}R_{\ l]}%
^{j]}+\frac{R}{\left(  n-1\right)  \left(  n-2\right)  }\delta_{kl}^{ij}
\label{dec}%
\end{equation}
reduces to%
\begin{equation}
C_{\ kl}^{ij}:=R_{\ kl}^{ij}-\gamma\delta_{kl}^{ij}\ . \label{rieem}%
\end{equation}
Here antisymmetrization is normalized as $T_{[ij]}:=\frac{1}{2}(T_{ij}%
-T_{ji})$. \newline

The Gauss-Bonnet tensor (\ref{egb2}) can then be expressed as%
\begin{equation}
H_{\ j}^{i}=C_{\ lm}^{ik}C_{\ jk}^{lm}-\frac{1}{4}\left[  C^{2}+\gamma
^{2}(n-1)(n-2)(n-3)(n-4)\right]  \delta_{j}^{i}, \label{g2n}%
\end{equation}
where $C^{2}:=C_{\ kl}^{ij}C_{\ ij}^{kl}$. Note that for Euclidean signature
$C^{2}\geq0$, and it vanishes only if $C_{\ kl}^{ij}=0$. Thus, by virtue of
(\ref{rieem}), Euclidean Einstein manifolds with $C^{2}=0$ are of constant
curvature.\newline

The trace of Eq. (\ref{g2n}) implies that the difference of the Gauss-Bonnet
combination and the squared Weyl tensor is a constant, i.e.,%
\begin{equation}
R_{kl}^{ij}R_{ij}^{kl}-4R^{ij}R_{ij}+R^{2}-C^{2}=\frac{n!}{\left(  n-4\right)
!}\gamma^{2}\ , \label{scr}%
\end{equation}
which is actually valid for $n>3$. Note that since in four dimensions the
Gauss-Bonnet tensor identically vanishes, $H_{\ j}^{i}\equiv0$, Eq.
(\ref{g2n}) means that Einstein manifolds fulfill the following identity
\cite{knt}%
\begin{equation}
C_{\ lm}^{ik}C_{\ jk}^{lm}=\frac{C^{2}}{4}\delta_{j}^{i}\ .
\end{equation}

Another useful identity allows writing Eq. (\ref{Trace-generic-D}) as:%
\begin{equation}
\mathcal{R}_{\ kl}^{ij}\mathcal{R}_{~ij}^{kl}-4\mathcal{R}_{ij}\mathcal{R}%
^{ij}+\mathcal{R}^{2}=0\ ,
\end{equation}
with%
\begin{equation}
\mathcal{R}_{\ kl}^{ij}:=\tilde{R}_{\ kl}^{ij}-\gamma\delta_{kl}^{ij}\ .
\end{equation}

\end{document}